\documentclass[11pt]{article}
\usepackage{amssymb,amsmath,mathrsfs,enumerate}
\usepackage{amsfonts, dsfont}
\usepackage{url}
\usepackage{graphicx,rotate,multicol}
\usepackage[margin=10pt,labelfont=bf]{caption}
\usepackage{cite}
\usepackage[utf8]{inputenc}
\usepackage[colorlinks=true,
            linkcolor=red,
            urlcolor=blue,
        citecolor=blue]{hyperref}

\usepackage{color}
\usepackage{lineno}

\def\ms{\mathscr}
\def\rt32{{\surd3 \over 2}}

\long\def\rpl#1!!!#2!!!{\color[rgb]{.7,0,0}{#1} \color{blue}{#2} \color{black}}

\let\tilde=\widetilde

\let\bar=\overline

\def \order(#1){{\mathcal O} \left(#1 \right)}
\def\rep#1{\ensuremath{\mathbf #1}}

\evensidemargin=\oddsidemargin
\def\Eqn#1{Eq.\ (\ref{#1})}
\def\Eqs#1#2{Eqs.\ (\ref{#1}) and (\ref{#2})}
\def\3Eqs#1#2#3{Eqs.\ (\ref{#1}), (\ref{#2}) and (\ref{#3})}
\def\sec#1{Sec.\,\ref{#1}}

\def\(({(\!(}
\def\)){)\!)}

\def\bsup#1{^{(#1)}}
\def\rep#1{\ensuremath{\mathit #1}}
\def\vev#1{\left< #1 \right>}

\allowdisplaybreaks

\title	{\LARGE\bf 
An $S_3$ flavored left-right symmetric model of quarks
}

\author {\sf Dipankar
  Das,\footnote{ddphy@caluniv.ac.in} \quad Palash
B. Pal\footnote{palashbaran.pal@saha.ac.in } \\[10pt]
\small\em Department of Physics, University of Calcutta, 92
Acharya Prafulla Chandra Road, Kolkata 700009, India\\ 
}

\date{}

\begin{document}


\maketitle	

\begin{abstract}
We construct a model based on the electroweak gauge group
$\rm SU(2)_L \times SU(2)_R \times U(1)_{B-L}$ augmented by
an $S_3$ symmetry. We assign nontrivial $S_3$ transformation
properties to the quarks and consequently we need two scalar
bidoublets. Despite the extra bidoublet we have only six Yukawa
couplings thanks to the discrete symmetry. Diagonalization of
the quark mass matrices shows that at the leading order only
the first two generations mix, resulting in a block diagonal CKM
matrix, and the first generation quarks are massless. Inclusion
of subleading terms produce an acceptable CKM matrix up to corrections of
$\order(\lambda^4)$. As for the first generation quark masses,
we obtain satisfactory value of $m_u/m_d$. The masses
themselves, though being in the same ballpark, 
 turn out to be somewhat smaller than the accepted
range.
\end{abstract}

\bigskip

\section{Introduction}\label{s:in}
One very compelling extension of the standard model (SM) is the
left-right symmetric (LRS) model\cite{Mohapatra:1974gc, 
Senjanovic:1975rk} based on the electroweak
gauge group $\rm SU(2)_L \times SU(2)_R \times U(1)_{B-L}$.  Unlike
the SM, the left-chiral and right-chiral fermions are treated
similarly in these models.

In the minimal LRS model, the left-chiral and right-chiral quarks are
assigned the following representations under the gauge group:
\begin{eqnarray}
Q_{iL} : (2,1, \textstyle{\frac13}) \,, \qquad
Q_{iR} : (1,2,\frac13) \,, 
\end{eqnarray}
where the index $i$ runs from 1 to 3 to accommodate three generations.
Allowing the Yukawa couplings for the quarks requires the presence 
of a scalar bidoublet
\begin{eqnarray}
\Phi : (2,2,0) \,.
\label{Phi}
\end{eqnarray}
It has two neutral components and therefore two possible
vacuum expectation values (VEVs).  The
quarks obtain their masses after symmetry breaking through the VEVs of
this $\Phi$ field.  The mass matrix has many free parameters.  
 There are nine Yukawa
couplings involving $\Phi$ that relate three generations of
left-chiral quarks with three generations of right-chiral quarks.
Besides, since the representation of $\Phi$ under the gauge group is
real, there are nine more couplings where $\Phi$ is replaced by its
complex conjugate,
\begin{eqnarray}
\tilde\Phi = \tau_2 \Phi^* \tau_2 \,,
\label{tildePhi}
\end{eqnarray}
suitably sandwiched by the antisymmetric Pauli matrix so that its
transformation property is exactly the same as that of $\Phi$.
Because of this large number of parameters, the quark mass matrices do
not have much of a predictive power.

In this article, we impose an $S_3$ symmetry between the generations
and show that the number of Yukawa couplings is drastically reduced to the extent
that predictions are possible.  Such a symmetry has been explored
extensively in the context of the SM gauge group\cite{Pakvasa:1977in, Deshpande:1991zh,
Koide:1999mx, Kubo:2003iw, Harrison:2003aw, Chen:2004rr, Koide:2006vs,
Chen:2007zj, Jora:2009gz, Kaneko:2010rx, Xing:2010iu, Teshima:2011wg, Dev:2011qy, Canales:2012dr, Canales:2013cga,
Dias:2012bh, Meloni:2012ci, Dev:2012ns, Zhou:2011nu, Ma:2013zca,
Benaoum:2013ji, Ma:2014qra, Cogollo:2016dsd, Pramanick:2016mdp,
Varzielas:2016zuo, Kubo:2004ps, Teshima:2005bk, Koide:2005ep, 
Bhattacharyya:2010hp, Bhattacharyya:2012ze, Teshima:2012cg, 
Barradas-Guevara:2014yoa, Hernandez:2015dga, Emmanuel-Costa:2016vej,
 Das:2014fea, Das:2015sca, Das:2017zrm}.
However, to our knowledge, this discrete symmetry in the context of
the left-right symmetric gauge group\cite{Garces:2018nar, Gomez-Izquierdo:2017rxi} has not been explored very much.
We will show that the enhanced
gauge symmetry, along with the discrete symmetry, leads to relations
between the quark masses and mixings.

\section{The model with a horizontal $S_3$ symmetry}\label{s:mh}
We extend the minimal LRS model with an extra $S_3$ symmetry that acts
between different generations of fermions.  This $S_3$ symmetry has
three different irreducible representation, \rep1, \rep{1'} and \rep2,
where the numbers signify the dimension of the representation
matrices.  The group has one order-2 and one order-3 generators.
In the \rep2 representation, we take them to be
\begin{eqnarray}
g_2 = \begin{bmatrix}\frac12 &
  \frac{\surd3}2  \\ \frac{\surd3}2 & - \frac12 
  \end{bmatrix}\,,
\qquad
g_3 = \begin{bmatrix} -\frac12 & \frac{\surd3}2 \\
- \frac{\surd3}2 & -\frac12
\end{bmatrix} \,, 
\label{e:gen}
\end{eqnarray}
For this choice of basis, we assign the following representations to
the fermions
\begin{eqnarray}
\begin{bmatrix} Q_1  \\ Q_2
  \end{bmatrix} : \rep2 \,, \qquad Q_3 : \rep1 \,,
\end{eqnarray}
following the same rule for left and right chiral quarks.  In order to
obtain an acceptable mass pattern, we now need scalars to be in the
\rep2 representation of $S_3$.  This means that we need to add an
extra bidoublet over and above what was shown in \Eqn{Phi}\cite{twobi}.
Calling the two scalar multiplets $\Phi_1$ and $\Phi_2$, we assign
them the representation
\begin{eqnarray}
\begin{bmatrix} \Phi_1  \\ \Phi_2  \end{bmatrix} : \rep2 
\end{eqnarray}
under the $S_3$ symmetry.  Keeping in mind
the fact that for any term in the Lagrangian where there is
a $\Phi$, there is another term containing $\tilde\Phi$, 
we can write down the most general Yukawa couplings
involving quarks as:
\begin{eqnarray}
- \ms L_Y &=& A \Big( \bar Q_{1L} \Phi_1 + \bar Q_{2L} \Phi_2
\Big) Q_{3R} + C \bar Q_{3L} \Big( \Phi_1 Q_{1R}
+ \Phi_2  Q_{2R} \Big) \nonumber\\*
&& + B \bigg[ \Big( \bar Q_{1L} \Phi_2 + \bar Q_{2L}
\Phi_1 \Big) Q_{1R} + \Big( \bar Q_{1L} \Phi_1 - \bar Q_{2L}
\Phi_2 \Big) Q_{2R} \bigg] \nonumber\\*
&&  + \tilde A \Big( \bar Q_{1L} \tilde\Phi_1 + \bar Q_{2L} \tilde\Phi_2
\Big) Q_{3R} + \tilde C \bar Q_{3L} \Big( \tilde\Phi_1 Q_{1R}
+ \tilde\Phi_2  Q_{2R} \Big) \nonumber\\*
&& + \tilde B \bigg[ \Big( \bar Q_{1L} \tilde\Phi_2 + \bar Q_{2L}
\tilde\Phi_1 \Big) Q_{1R} + \Big( \bar Q_{1L} \tilde\Phi_1 - \bar Q_{2L}
\tilde\Phi_2 \Big) Q_{2R} \bigg] + \mbox{h.c.} \,.
\end{eqnarray}
After symmetry breaking, both $\Phi_1$ and $\Phi_2$ develop VEVs:
\begin{eqnarray}
\vev {\Phi_a} = \begin{pmatrix} \kappa_a & 0 \\ 0 & \kappa'_a
\end{pmatrix} \,, \qquad a=1,2.
\end{eqnarray}
The resulting mass matrices for the quarks are of the form
\begin{subequations}
	\label{mh.M}
	\begin{eqnarray}
	\ms M_u = F \kappa_1 + G \kappa_2 + \tilde F \kappa_1' +
	\tilde G \kappa_2' \,,
	\label{mh.Mu} \\
	\ms M_d = \tilde F \kappa_1 + \tilde G \kappa_2 + F \kappa_1' +
	G \kappa_2' \,, 
	\label{mh.Md}
	\end{eqnarray}
\end{subequations}
where
\begin{eqnarray}
F = \begin{pmatrix}
0 & B & A \\
B & 0 & 0 \\
C & 0 & 0
\end{pmatrix} \,, \qquad 
G = \begin{pmatrix}
B & 0 & 0 \\
0 & -B & A \\
0 & C & 0
\end{pmatrix} \,,
\end{eqnarray}
and $\tilde F$ and $\tilde G$ are matrices which have exactly the same
form, except that they involve the Yukawa couplings with tilde marks.
We will assume that all Yukawa couplings are real, and so are the
VEVs.  Our task is now to perform the diagonalization of the mass
matrices given in \Eqn{mh.M} and show that, under some reasonable
assumptions, the diagonalization can be performed and the quark mixing
matrix can be obtained in a form that is consistent with the present
data.

Of course the matrices shown in \Eqn{mh.M} cannot be diagonalized in
general with the help of unitary transformations.  One needs bi-unitary
transformations, which induces different transformations on the
left-chiral and right-chiral quarks.  For the sake of the CKM matrix,
we need only the mixing of the left-chiral fermions.  The relevant
mixing matrices can be obtained by considering the diagonalization of
$\ms M_q\ms M_q^\dagger$, where the index $q$ takes two values, $u$
and $d$, to distinguish the up-sector quarks from the down-sector
quarks.  Let us write
\begin{eqnarray}
U_q \ms M_q \ms M_q^\dagger U_q^\dagger = D_q^2 \,,
\label{mh.Dsq}
\end{eqnarray}
where $D_q^2$ is a diagonal matrix whose diagonal elements are the
mass-squared values of the quarks of type $q$ (i.e., $u$ or $d$).
Then the CKM matrix will be given by
\begin{eqnarray}
V_{\rm CKM} = U_{u\vphantom d}^{\phantom\dagger} U_d^\dagger \,.
\label{mh.ckm}
\end{eqnarray}
We therefore need to find the diagonalizing matrices $U_u$ and $U_d$.
For this, we need to proceed in steps, making some assumptions which
we now describe.

\section{Large and small terms}\label{s:ls}
In order to perform the diagonalization, we will first make some
assumptions about the relative magnitudes of different parameters.
The first thing we assume is that the primed VEVs are much smaller
compared to the unprimed ones:
\begin{eqnarray}
\kappa_1', \kappa_2' \ll \kappa_1, \kappa_2 \,.
\end{eqnarray}
The opposite assumption $\kappa_1', \kappa_2' \gg \kappa_1, \kappa_2$
will serve as well, and amounts to fixing a convention.
Such an assumption can naturally suppress the mixing between
the  gauge bosons in the left and right sectors.
The terms in
\Eqn{mh.M} proportional to the unprimed VEVs will therefore be
considered dominant, and the other terms, proportional to the primed
VEVs, will be considered as perturbations.  In this section, we
consider diagonalization of the quark mass matrices in the limit
$\kappa_1'=\kappa_2'=0$, {\it i.e.}, in the zeroth order of smallness.

There are only two VEVs at this level of approximation, $\kappa_1$ and
$\kappa_2$.  Since the other VEVs have been assumed to be negligible,
we can write
\begin{eqnarray}
\kappa_1^2 + \kappa_2^2 = v^2 \,,
\label{0a.v}
\end{eqnarray}
where $v=174$~GeV is the breaking scale of the SM.  We also define
\begin{eqnarray}
\tan \beta = {\kappa_2 \over \kappa_1} \,.
\end{eqnarray}
Henceforth, instead of using $\kappa_1$ and $\kappa_2$ directly, we
will use the parameters $v$ and $\beta$.

Thus, in this zeroth order approximation, the mass matrices of the quarks are of
the form
\begin{eqnarray}
\ms M_q \bsup0 = v \begin{pmatrix}
B_q \sin\beta & B_q \cos\beta & A_q \cos\beta \\
B_q \cos\beta & -B_q \sin\beta & A_q \sin\beta \\
C_q \cos\beta & C_q \sin\beta & 0 
\end{pmatrix} \,,
\label{0a.M0}
\end{eqnarray}
where, for the ease of notation, we have renamed the Yukawa couplings
by a subscript $q$  according to which mass matrix they
contribute to:
\begin{subequations}
	\label{0a.Y_q}
	\begin{eqnarray}
	A_u = A \,, \qquad & B_u = B \,, & \qquad C_u = C \,,
	\label{0a.Y_u} \\
	A_d = \tilde A \,, \qquad & B_d = \tilde B \,, & \qquad C_d = \tilde
	C \,. 
	\label{0a.Y_d}
	\end{eqnarray}
\end{subequations}
The kind of mass matrix shown in \Eqn{0a.M0} was obtained in our
earlier work\cite{Das:2017zrm} in the context of $\rm SU(2)_L \times
U(1)_Y$ model.  In order to perform a diagonalization of the mass
matrices at this level of approximation, we note that
%
\begin{eqnarray}
|\det (\ms M_q \bsup0)| = v^3 A_q B_q C_q \sin 3\beta \,.
\label{e:det0}
\end{eqnarray}
Since the first generation quark masses are very small, we assume
that they are zero at this level, and arise entirely
from smaller corrections to the mass matrices.  Then the determinant
must vanish at this level.  Without arbitrarily making some of the
Yukawa couplings vanish, this can be achieved, in both up and down
sectors, if we have
\begin{eqnarray}
\sin 3\beta = 0 \,.
\label{e:3b0}
\end{eqnarray}
This value can be nontrivially obtained by setting
\begin{eqnarray}
\beta = \pi/3 \,.
\label{0a.beta}
\end{eqnarray}
We assume that this is indeed the value of $\beta$ that comes out of
the minimization of the Higgs potential at this level of
approximation, i.e., on assuming $\kappa_1' = \kappa_2' = 0$.  Then,
taking
\begin{eqnarray}
U_q \bsup0 = \begin{pmatrix}
- {\surd3 \over 2} \sin\theta_q & \frac12 \sin\theta_q &
\cos\theta_q \\
{\surd3 \over 2} \cos\theta_q & - \frac12 \cos\theta_q &
\sin\theta_q \\
\frac12 & {\surd3 \over 2} & 0 
\end{pmatrix} 
\label{0a.Uq0}
\end{eqnarray}
with
\begin{eqnarray}
\tan\theta_q = {C_q \over B_q} \,,
\label{0a.thq}
\end{eqnarray}
one finds
\begin{eqnarray}
U_q \bsup0 \ms M_q \bsup0 {\ms M_q \bsup0}^\dagger  {U_q \bsup0}^\dagger
= v^2 \begin{pmatrix}
0 & 0 & 0 \\
0 & B_q^2+C_q^2 & 0 \\
0 & 0 & A_q^2+B_q^2 
\end{pmatrix} \,.
\label{0a.UMMU}
\end{eqnarray}
At this stage, then, the CKM matrix is given by
\begin{eqnarray}
V_{\rm CKM} \bsup0 =  U_u \bsup0  {U_d \bsup0}^\dagger
= \begin{pmatrix}
\cos(\theta_u-\theta_d) & - \sin(\theta_u-\theta_d) & 0 \\
\sin(\theta_u-\theta_d) & \cos(\theta_u-\theta_d) & 0 \\
0 & 0 & 1 
\end{pmatrix} \,.
\label{e:CKM0}
\end{eqnarray}
This shows that at the zeroth order, we have only the Cabibbo angle that
mixes the first two generations of quarks, whereas the third
generation is unmixed.  This state of affairs is certainly consistent
with the fact that the Cabibbo angle in the largest angle in the CKM
matrix, and all others are much smaller.  In \sec{s:st}, we will see
how the small angles can arise from the small corrections that we
have  left out so far.

Before that, we want to summarize the information that we have already
obtained about the masses and consequently about the Yukawa couplings.
From \Eqn{0a.UMMU}, we see that at the zeroth level of approximation,
\begin{subequations}
	\label{0a.tcbs}
	\begin{eqnarray}
	m_t^2 = (A_u^2+B_u^2) v^2 \,, &\qquad& m_c^2 = (B_u^2+C_u^2) v^2 \,,
	\label{0a.tc} \\ 
	m_b^2 = (A_d^2+B_d^2) v^2 \,, &\qquad& m_s^2 = (B_d^2+C_d^2) v^2 \,.
	\label{0a.bs}
	\end{eqnarray}
\end{subequations}
Although these masses will receive some corrections which will be introduced
later, such modifications are expected to be small, and therefore we can use
\Eqn{0a.tcbs} as a very good approximation to the actual masses.
Knowledge of the hierarchy of quark masses then tells us that
\begin{eqnarray}
A_u^2 \gg B_u^2 , C_u^2 \,,  \qquad
A_d^2 \gg B_d^2 , C_d^2 \,, 
\label{0a.A>>}
\end{eqnarray}
so that the third generation is much heavier than the second, and
further
\begin{eqnarray}
A_u^2 \gg A_d^2
\label{0a.Au>>Ad}
\end{eqnarray}
to ensure that the top mass is much bigger than the bottom mass.
Using  \Eqn{0a.A>>} and the definition of
\Eqn{0a.thq}, we can write the Yukawa couplings in the form
\begin{eqnarray}
A_q \approx {m_{3q} \over v} \,, \qquad B_q \approx {m_{2q} \over v}
\cos\theta_q \,, 
\qquad C_q \approx {m_{2q} \over v} \sin\theta_q \,,  
\label{e:23gen}
\end{eqnarray}
where $m_{3q}$ and $m_{2q}$ denote the masses of the third and second
generation quarks in the sector marked by $q$, {\it i.e.},
\begin{eqnarray}
m_{3u} = m_t \,, \qquad m_{2u} = m_c \,, \qquad m_{3d} = m_b \,,
\qquad m_{2d} = m_s \,. 
\end{eqnarray}

At this point, perhaps it is worth reemphasizing the main
conclusion of this section. Here we have considered an approximate
reality where $m_u=m_d=0$ and $V_{i3}=V_{3i}=0$ ($i=1,2$) as well.
The vanishing of $V_{3i}$ ($i=1,2$) will follow automatically from the
vanishing of $V_{i3}$ ($i=1,2$) due to the unitarity of the CKM
matrix.  We are thus left with four zeros, viz., $m_u=0$, $m_d=0$ and
$V_{i3}=0$ ($i=1,2$), which are disconnected in the SM, i.e., they are
four different {\em accidents} in the framework of the SM. But, in our
model, one needs only one {\em accident}, given by \Eqn{e:3b0}, to
achieve all these zeros, {\it i.e.}, the four zeros are
connected. Therefore, concerning the small parameters in the quark
Yukawa sector, our construction provides a sense of aesthetic
connection that is absent in the SM.
Moreover,  this approximate reality with $\kappa'_a=0$ forbids
$W_L$-$W_R$ mixing. In the next section, we will see that turning 
on small values of $\kappa'_a$ leads to small CKM elements as
well as the first generation quark masses. Therefore, in our scenario,
these small masses and mixings in the quark sector owe their origin
to the same parameters which govern
the smallness of the $W_L$-$W_R$ mixing.

\section{Including the smaller terms}\label{s:st}
We now try to see the effects of non-zero values of $\kappa_1'$ and
$\kappa_2'$.  The extra contributions that appear in the mass matrices
will be denoted by $\ms M'$, {\it i.e.},
\begin{eqnarray}
\ms M_q = \ms M_q \bsup0 + \ms M'_q \,.
\end{eqnarray}
These contributions will come from two
sources.  First, there are terms proportional to $\kappa_1'$ and
$\kappa_2'$ in \Eqn{mh.M}.  Second, the minimization of the Higgs
potential will now not give \Eqn{0a.beta}, but rather
\begin{eqnarray}
\sin 3\beta = 3 \delta \,,
\end{eqnarray}
with some small value of $\delta$.

All correction terms in the mass matrices will have one factor of some
Yukawa coupling.  Motivated by the hierarchy among the Yukawa
couplings noted in \sec{s:ls}, we will keep only the terms
proportional to $A_u$ as the dominant corrections to $\ms M_q \bsup0$
defined in \Eqn{0a.M0}, with the understanding that the
contribution from other terms are proportional to much smaller 
Yukawa couplings, and are negligible at the level of accuracy that 
we seek for.  Keeping these in mind, we can write the dominant
 corrections are as follows:
\begin{eqnarray}
\ms M'_u \approx vA_u \begin{pmatrix}
0 & 0 &  \rt32 \delta \\
0 & 0 & -\frac12 \delta \\
0 & 0 & 0 \\
\end{pmatrix} \,, \qquad
\ms M'_d \approx A_u \begin{pmatrix}
0 & 0 & \kappa_1' \\
0 & 0 & \kappa_2' \\
0 & 0 & 0 \\
\end{pmatrix}  
\label{st.M'}
\end{eqnarray}
In order to set up a uniform notation for both up and down sectors, let
us introduce some shorthands through the relations
\begin{subequations}
	\begin{eqnarray}
	\Big( \ms M'_u \Big)_{13} = m_t \epsilon_u \cos \chi_u \,, &\qquad&
	\Big( \ms M'_u \Big)_{23} = m_t \epsilon_u \sin \chi_u \,, \\
	\Big( \ms M'_d \Big)_{13} = m_t \epsilon_d \cos \chi_d \,, &\qquad&
	\Big( \ms M'_d \Big)_{23} = m_t \epsilon_d \sin \chi_d \,, 
	\end{eqnarray}
\end{subequations}
so that 
\begin{subequations}
	\begin{eqnarray}
	\epsilon_u = \delta \,, &\qquad& \chi_u = - \pi/6 \,, \\ 
	\epsilon_d = \sqrt{\kappa_1'^2+\kappa_2'^2} \Big/v \,, &\qquad&
	\tan\chi_d = \kappa_2' / \kappa_1' \,.
	\end{eqnarray}
\end{subequations}
We now need to examine the mass matrices including these corrections, and
the diagonalization procedure.

The first thing that we notice is that, after the inclusion of $\ms
M'$, the determinant of the mass matrix is no more zero and
is given by
\begin{eqnarray}
|\det \ms M_q| = v^2 B_q C_q m_t \epsilon_q \sin \Big( \frac\pi3 - \chi_q
\Big) \,.
\end{eqnarray}
This quantity should be equal to the product of the three mass
eigenvalues. Therefore, the mass of the first generation quark
will be given by
\begin{eqnarray}
m_{1q} = {m_t m_{2q} \over m_{3q}} \sin \theta_q \cos\theta_q 
\epsilon_q \sin \Big( \frac\pi3 - \chi_q \Big)
=\epsilon'_q m_{2q} \sin\theta_q \cos\theta_q \,,
\label{st.m1q} 
\end{eqnarray}
where we have substituted $B_q$ and $C_q$ using \Eqn{e:23gen}
and defined
\begin{eqnarray}
\epsilon'_q = \frac{m_t}{m_{3q}}\epsilon_q \sin \Big( \frac\pi3 - \chi_q \Big) \,.
\label{e:epqp}
\end{eqnarray}
In a less cluttered but lengthier way, we can break up \Eqn{st.m1q} as
\begin{subequations}
	\label{e:mumd}
	\begin{eqnarray}
	m_u &=& \epsilon'_u m_c \sin \theta_u \cos\theta_u  \,, \\
	m_d &=& \epsilon'_d m_s \sin \theta_d \cos\theta_d \,.
	\end{eqnarray}
\end{subequations}

We now look at the diagonalization of the matrices $\ms M_q \ms
M_q^\dagger$.  Referring back to \Eqn{mh.Dsq} and its zeroth level
analog, \Eqn{0a.UMMU}, we propose to incorporate the correction to the
diagonalizing matrix by writing
\begin{eqnarray}
U_q = X_q U_q \bsup0 \,,
\label{st.Uq}
\end{eqnarray}
where $X_q$ is supposed to inflict small corrections on $U_q \bsup0$.
We now parametrize $X_q$ by writing
\begin{eqnarray}
X_q = \begin{pmatrix}
1  & 0 & \alpha_q \\
0 & 1  & \gamma_q \\
-\alpha_q & - \gamma_q & 1
\end{pmatrix} \,,
\label{e:Xq}
\end{eqnarray}
ignoring higher order terms in $\alpha_q$ and $\gamma_q$.
We have checked that including a rotation in the $12$ sector as well
contributes only at a subleading order. Therefore,
for our purposes, \Eqn{e:Xq} constitutes a reasonable
approximation for the correction to $U_q \bsup0$.

If we now evaluate
the left side of \Eqn{mh.Dsq}, using \Eqn{st.Uq} for $U_q$ and $\ms
M_q$ as the sum of the expression of \Eqn{0a.M0} with $\beta=\pi/3$
and the corrections from \Eqn{st.M'}, we should obtain a diagonal
matrix, to the accuracy employed in defining the small parameters.
From this condition, one should be able to determine the relevant
parameters of $X_q$.  

First, we check the diagonal elements.  The lower two diagonal
elements will pick up small corrections to the formulas of the second and
third generation quarks given in \Eqn{0a.tcbs}, 
and are unimportant for our purpose.  The
first diagonal element should give the mass squared of the first
generation quark.  Evaluation of \Eqn{mh.Dsq} gives
\begin{eqnarray}
m_{1q}^2 &=& (m_{2q}^2 \cos^2\theta_q + m_{3q}^2) \alpha_q^2 - 2 m_t
m_{3q} \epsilon_q \sin \Big(
\frac\pi3 - \chi_d \Big) \alpha_q \nonumber\\*
&& + m_t^2 \epsilon_q^2 \sin^2 \theta_q \sin^2 \Big( \frac\pi3 -
\chi_d \Big) \,.
\end{eqnarray}
But the mass value has already been found in \Eqn{st.m1q} from the
consideration of the determinant.  Putting in the value from there and
neglecting terms which provide corrections of order
$m_{2q}^2/m_{3q}^2$, we can determine $\alpha_q$ as:
\begin{eqnarray}
\alpha_q = {m_t \over m_{3q}} \; \epsilon_q \sin \theta_q \sin \Big(
\frac\pi3 - \chi_q \Big) \equiv \epsilon'_q \sin\theta_q \,.
\label{e:aq}
\end{eqnarray}
Quite nicely, the 13 element of the left side of
\Eqn{mh.Dsq} also vanishes at the leading order under the same condition, 
confirming the consistency of the approximation.  Further, the vanishing of the 23 element at the leading order gives the expression for $\gamma_q$ as
\begin{eqnarray}
\gamma_q = -{m_t \over m_{3q}} \; \epsilon_q \cos \theta_q \sin \Big(
\frac\pi3 - \chi_q \Big) \equiv -\epsilon'_q \cos\theta_q \,.
\label{e:gq}
\end{eqnarray}

\begin{figure}[htbp]
\centering
\includegraphics[scale=0.99]{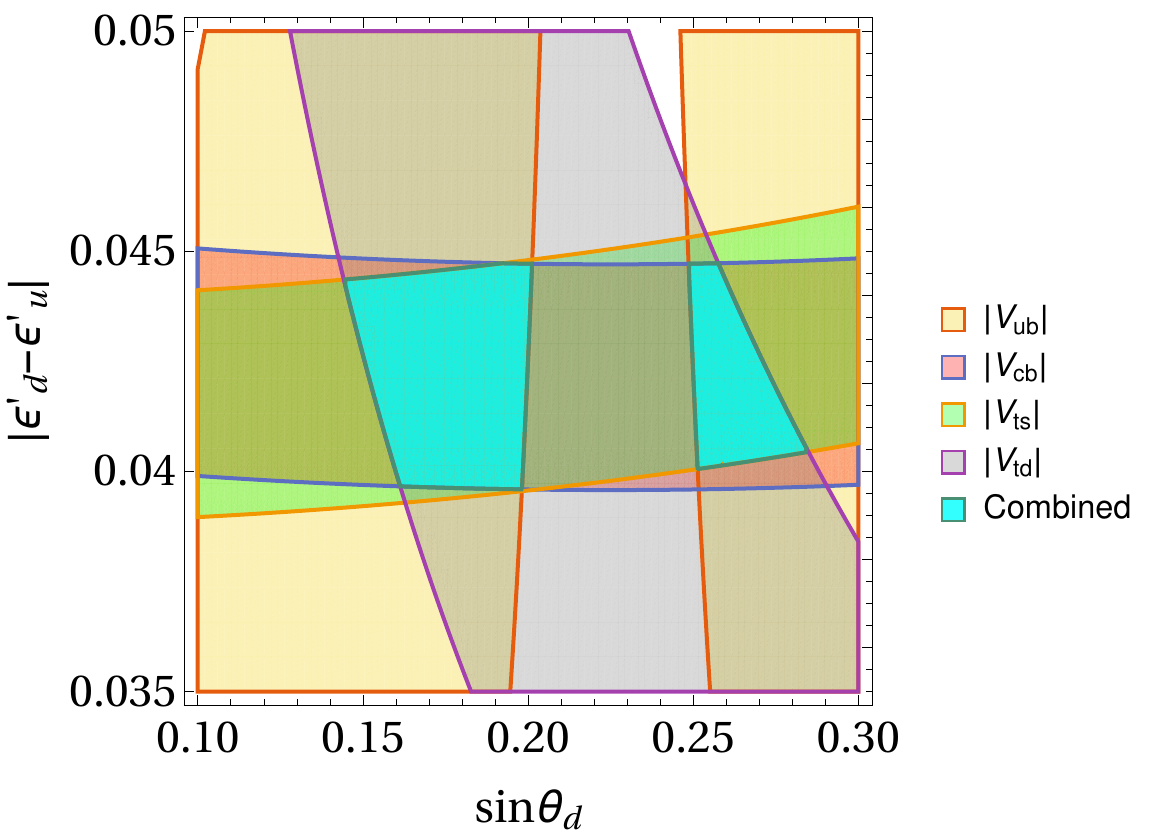}
\caption{\em Shaded areas in  yellow,  orange, green and gray represent allowed
regions from $|V_{ub}|$, $|V_{cb}|$, $|V_{ts}|$ and $|V_{td}|$ respectively.
The area shaded in cyan represents the common solution region.}
\label{f:fit}
\end{figure}

\section{The CKM matrix}\label{s:ckm}
Using \Eqn{mh.ckm} in conjunction with \Eqn{st.Uq}, we can now write the
CKM matrix as
\begin{eqnarray}
V_{\rm CKM} = X_u U_u \bsup0 {U_d \bsup0}^\dagger X_d^\dagger &=& X_u V_{\rm
	CKM} \bsup0 X_d^\dagger \nonumber\\ 
	&\approx& \begin{pmatrix}
	\cos\theta_C  &  -\sin\theta_C  &  -(\epsilon'_d-\epsilon'_u)\sin\theta_u  \\
	\sin\theta_C  &  \cos\theta_C  &  (\epsilon'_d-\epsilon'_u)\cos\theta_u   \\
	(\epsilon'_d-\epsilon'_u)\sin\theta_d  &  -(\epsilon'_d-\epsilon'_u)\cos\theta_d  &  1
	\end{pmatrix} \,,
\label{e:CKM}
\end{eqnarray}
where $V_{\rm CKM} \bsup0$ has been defined already in \Eqn{e:CKM0} and
the Cabibbo angle, $\theta_C$, is defined as
\begin{eqnarray}
\theta_C = \theta_u - \theta_d  \,.
\label{e:tC}
\end{eqnarray}
In writing \Eqn{e:CKM} we have also used the definition of $X_q$ given
in \Eqn{e:Xq} along with the solutions of \Eqs{e:aq}{e:gq}.

In the Wolfenstein parametrization \cite{Wolfenstein:1983yz} of the
CKM matrix, the off-diagonal 12 and 21 elements are of
$\order(\lambda)$, where $\lambda$ is a small parameter that is
roughly equal to the Cabibbo angle.  The 23 and 32 elements are
$\order(\lambda^2)$, whereas the 13 and 31 elements are
$\order(\lambda^3)$.  Since we have already produced the Cabibbo
mixing of $\order(\lambda)$ at the zeroth order, the perturbations
$\epsilon'_q$ should be at least of $\order(\lambda^2)$.
Taking $\epsilon'_q \sim \order(\lambda^2)$ and $\sin\theta_q \sim
\order(\lambda)$, we can see that \Eqn{e:CKM} reproduces the correct
orders of magnitudes for the different CKM elements.  For easy
comparison, we summarize below the current experimental values for the
magnitudes of the elements of the CKM matrix\cite{Koppenburg:2017mad,
  PDG}:
\begin{eqnarray}
|V_{\rm CKM}^{\rm exp}| =  \begin{pmatrix}
	0.97446\pm 0.00010 & 0.22452\pm 0.00044 & 0.00365\pm 0.00012  \\
	0.22438\pm 0.00044 & 0.97359\pm 0.00011 & 0.04214\pm 0.00076  \\
	0.00896\pm 0.00024 & 0.04133\pm 0.00074 & 0.999105\pm 0.000032
	\end{pmatrix}\,.
\label{e:CKMexp}
\end{eqnarray}
While comparing with the experimental values, we should keep in mind
that the inherent uncertainty of $\order(\lambda^4)$ in \Eqn{e:CKM} is
much larger than the experimental uncertainties in
\Eqn{e:CKMexp}. Therefore, for the $ij$-th element of the CKM matrix,
we take
\begin{eqnarray}
V_{ij} = V_{ij}^{\rm cen} \pm \lambda^4 \,,
\label{e:Vij}
\end{eqnarray}
where the central values are taken from \Eqn{e:CKMexp}. We assume that
$\sin\theta_C \equiv -\lambda \approx -0.225$ has been measured quite
accurately and use \Eqn{e:tC} to express $\theta_u$ in terms of
$\theta_d$. Our goal is to see, using \Eqn{e:Vij}, whether there
exists a common region in the $\sin\theta_d$
vs. $|\epsilon'_d-\epsilon'_u|$ plane, which is allowed by $|V_{ub}|$,
$|V_{cb}|$, $|V_{ts}|$ and $|V_{td}|$ simultaneously. We display our
result in Fig.~\ref{f:fit} where we see that there is indeed some
common solution region.
Note that there are two different allowed regions from $|V_{ub}|$,
which correspond to different signs for $\sin\theta_u$.

With $|\epsilon'_d-\epsilon'_u|$ and $\sin\theta_d$ nearly fixed from
Fig.~\ref{f:fit}, now we have only one parameter, namely $\epsilon'_u$
(or equivalently $\epsilon'_d$) to play around. Thus, using
\Eqn{e:mumd}, we still need to reproduce {\em two} light quark masses,
$m_u$ and $m_d$, with only {\em one} parameter remaining at our
disposal. As a matter of fact, the cyan region on the left in
Fig.~\ref{f:fit}, is disfavored because it gives too small values for
the down-quark mass. Keeping this in mind, we choose the following
values
\begin{eqnarray}
	\epsilon'_d = 0.072   \,, \qquad  \epsilon'_u = 0.028    \,, \qquad
	\sin\theta_d = 0.26    \,,
\label{e:bench}	
\end{eqnarray}
which correspond to a benchmark point somewhere in the cyan region on
the right in Fig.~\ref{f:fit}. Using these values we find 
\begin{subequations}
\label{e:findings}
\begin{eqnarray}
  |V_{ub}|\approx 0.002 \,,  && \qquad  |V_{cb}|\approx 0.044  \,, \qquad
|V_{td}|\approx 0.011     \,,  \qquad  |V_{ts}|\approx 0.042   \,.
\end{eqnarray}
We see that these values of the CKM elements are acceptable within an
error bar of $\order(\lambda^4)$.  As commented earlier, we also
obtain the light quark masses from this exercise.
Taking $m_s=110$~MeV and $m_c=1.2$~GeV, and the values
of the parameters in \Eqn{e:bench}, we obtain using \Eqn{e:mumd}
\begin{eqnarray}
  m_u \approx 1.2 ~{\rm MeV} \,,   \qquad  m_d \approx 2.0 ~{\rm
  MeV} \,. 
\end{eqnarray}
\end{subequations}
The values of $m_u$ as well as
the ratio $m_u/m_d$ are within tolerable ranges, but the absolute
value of $m_d$ comes out to be a bit too low. Having only one free
remaining parameter prevents us
from obtaining a good fit for both the up and down quark masses. We
have checked that, as long as the down quark mass is concerned,
the values given in \Eqn{e:findings} reflects the best case scenario.

\section{Summary}
We have considered a model where the  left-right symmetric gauge group
$\rm SU(2)_L \times SU(2)_R \times U(1)_{B-L}$ is augmented by an
$S_3$ symmetry.  The discrete symmetry drastically reduces
the number of Yukawa couplings in the model.  In fact, there are only
six Yukawa couplings.  Because of the small number of parameters, we
can relate many aspects of quark masses and mixings satisfactorily
in our model.  We have demonstrated that the smallness of the first
generation quark masses is related to the smallness of the 13 and 23
elements of the CKM matrix as well as to the smallness of the $W_L$-$W_R$
mixing.  We have also shown that, under some
reasonable assumptions about the relative magnitudes of the VEVs, the
CKM matrix can be reproduced within an accuracy of $\order(\lambda^4)$.  The
only sore point seems to be the light quark masses that we obtain from
the model, which, although being in the same ballpark, 
turn out to be a bit smaller than expected. 
Yet the model deserves careful
attention since we believe that the successes
of the model with the CKM matrix outweigh the dissatisfaction with the
light quark masses.


\bibliographystyle{JHEP} 
\bibliography{s3_LRS.bib}

\end{document}